\documentstyle[11pt,newpasp,twoside,epsf]{article}
\def\edcomment#1{\iffalse\marginpar{\raggedright\sl#1\/}\else\relax\fi}
\reversemarginpar 

\begin{document}
\title{Dark Matter in the center of galaxies and galaxy clusters: ruling out
the CDM scenario?}
 \author{E. D'Onghia$^{1,2}$, G. Chincarini$^{2,3}$, C. Firmani$^{2}$, D. Marchesini$^{1}$}
\affil{1-Universita' degli Studi di Milano, Via Celoria 16, 
 20133 Milan, Italy, 2-Osservatorio Astronomico di Brera, Via E. Bianchi 46, 
23807 Merate, Italy, 3-Universita' degli Studi di Milano-Bicocca,
 Piazza dell'Ateneo Nuovo 1, 20126 Milan, Italy.}

\begin{abstract}
This work is focused on the preliminary results of the observations of 
H$\alpha$ rotation curves for some of the late type dwarf and LSB galaxies
carried out at the TNG telescope.
In light of the observational data and of the N-body simulations we have
performed recently, we discuss 
some of the implications on the nature of the DM particles and the 
formation of the dark haloes.
\vskip -1.1truecm
\end{abstract}

\section{Introduction}
Cold Dark Matter (CDM) models provide
a solid framework capable to explain most of the properties
of the universe at large scales. 
In the last few years a very large effort has been spent in the investigation
of the detailed structure of the virialized haloes within
the CDM scenario, in which dark particles are assumed to be cold and 
collisionless.
 Using N-body simulations it was showed that virialized 
haloes are well described by an universal density profile, the Navarro, 
Frenk, $\&$ White model (1997; hereafter NFW). This profile diverges at the 
center showing a cuspy core: $\rho \propto 1/r$. 
Recent high-resolution N-body simulations show that as the numerical
resolution is increased, the resulting dark density profile goes as  
$\rho \propto 1/r^{1.5}$ 
(Moore et al. 1999).
On galactic scales the theory seem 
to be in conflict with the observations, since the 
rotation curves of LSB and dwarf galaxies rise too slowly with respect to 
the predictions of the CDM model. Indeed, the rotation curves of these 
galaxies call for a finite central density (soft core), 
in conflict with the cuspy cores predicted by the model.
Note that  LSB and dwarf galaxies are systems
strongly dark matter dominated, so their rotation curves are good tracers
of the underlying dark haloes gravitational potential and therefore 
good candidates
to explore the innermost shape of the dark matter distribution.
Recently, however, some authors have challenged the existence of 
soft cores
in centers of galaxies measuring H$\alpha$ rotation curves of late-type
dwarf and LSB galaxies. Basically they find two results: HI rotation curves
for most of the galaxies are affected by beam-smearing and a 
good agreement with concentration values predicted by the NFW model (van
den Bosch and Swaters 2000). 

\section{Soft core observational evidence}
We have carried out H$\alpha$ rotation curves at the TNG of dwarf and LSB
galaxies. The spatial resolution in the central regions is a 
factor 10 better than with respect the HI data published in literature
for the same galaxies. 
In Fig.1 we reproduce the circular velocities for the galaxies 
UGC4325, UGC11861
and F571-8 (empty circles are our H$\alpha$ observations and filled
circles are HI data). From a comparison between optical and radio data 
it is interesting to note that HI observations for these galaxies are not 
affected (or are very little) by beam smearing (see Marchesini et al.
at this conference, for details).
To discriminate between the NFW and King models
as representative profile of the primeval  dark halo,  
 we built a fiducial galaxy. 
The dotted lines in the top panels represent a NFW profile
while the dotted lines in the bottom panels are due to a King profile.
Accounting for the disc formation (dashed lines)  we 
computed next the adiabatic contraction of
the DM component due to the cooling of the
baryons into the virialized haloes. 
The final rotation curve is then obtained (solid lines) and compared with 
the H$\alpha$ data.
The final curve is best fitted for all galaxies assuming that the 
dark matter in the primeval halo is distribuited as a King profile.
On galaxy cluster scale, there are at least two clusters with
evidence of soft cores: CL0024+1654 as inferred by strong lensing technique
(Tyson et al. 1998) and A1705 as revealed by the Chandra data 
(Ettori et al. at this conference).
The central densities of the galaxies analyzed here
or taken from the literature and 
of CL0024+1654 seem to confirm the evidence that the central density
is independent on the halo mass (Firmani et al. 2000).
Future observations will  
increase the sample and put these observational 
findings on a more robust basis.
\vskip -0.5truecm
\begin{figure}[h]   
\plotone{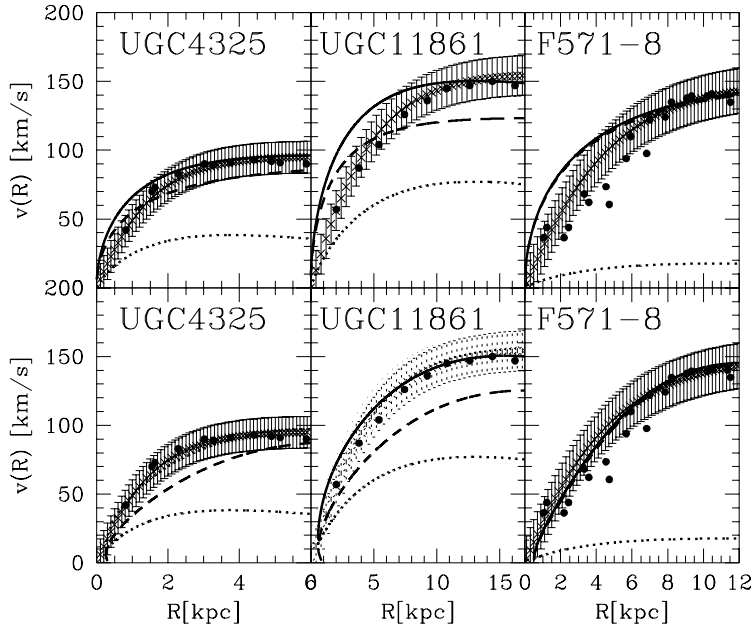}
\vskip -3.6truecm
\caption{Rotation curves of dwarf galaxies and LSB galaxies. 
{\it Top panels}: our H$\alpha$ observations with error bars
(skeletal), HI data
(filled points).
An initial NFW profile is assumed for the halo (dashed line),
the disc (dotted line) and the final rotation curve to be compared with data 
(solid line).{\it Bottom panels}: 
as above, however, an initial King profile is assumed for the dark
halo (dashed line).}
\vskip -0.5truecm
\end{figure}
\begin{figure}[h] 
\plotone{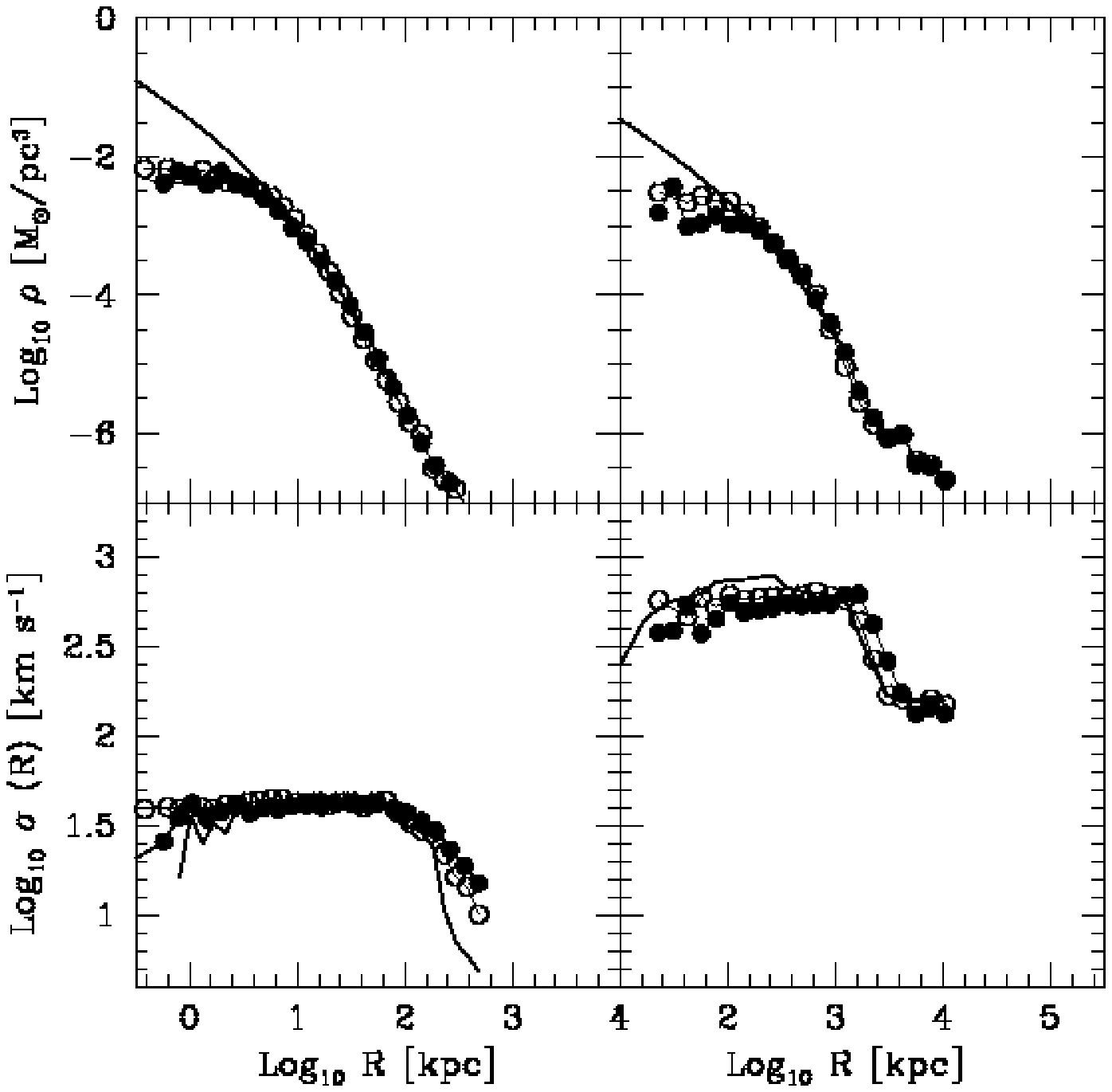}
\vskip -3.6truecm
\caption{{\it Top panels}. Dark density profiles in a SIDM scenario 
as derived by N-body simulations. {\it Bottom panels}. The corresponding
radial dispersion velocity profiles (see the text for details).}
\vskip -0.5truecm
\end{figure}

\section{Implications on the DM nature}
The question is: how can we justify the existence of soft cores in
galaxies and in galaxy clusters?
Warm dark matter (WDM) has been proposed in order to solve the conflict
with CDM models. 
However, even if interesting, this scenario is unable to
produce soft cores at galactic scales (see e.g.  
Avila-Reese et al. 2001). 
A possible solution is to assume a weakly self-interacting cold dark matter 
(SIDM)
(Spergel $\&$ Steinhardt 2000). It is well known that the NFW 
profile is far from  
thermal equilibrium, and shows, in fact, a inner positive temperture
gradient. Collisions between dark particles would modify
the velocity 
distribution to a Maxwellian distribution generating a  
constant velocity dispersion.
We have developed different approaches to explore the SIDM regime:
thermodynamic (Firmani et al. 2000) and dynamic models 
building a cosmological Boltzmann code (Firmani, D'Onghia, $\&$ Chincarini 
2000, astro-ph/0010497). Here, we propose a new numerical procedure applied to
N-body simulations (D'Onghia, Firmani, $\&$ Chincarini 2001, in preparation).
Our technique is capable to simulate a single halo of any size and mass
in a cosmological framework coupling semi-numerical technique and 
N-body simulations. The cosmological initial conditions are fixed by the mass
aggregation history of the halo using semi-analytical models and
only the dynamical evolution of the halo is followed by the N-body code.
To achieve this, the HYDRA code has been ad hoc modified
and our version can describe simulations both of collisional and collisionless
dark haloes. 
In our simulations we 
assume a cross section inversely proportional to the particle 
velocity dispersion: $\sigma/m_x v_{100}=10^{-24}$ cm$^2$/Gev
with $v_{100}$ the halo velocity dispersion in units of 100 km/s.    
The top panels of Fig. 2 show the dark density profiles of haloes of 
$M=10^{11} \ M_{\odot}$ (left panel) and $10^{15} \ M_{\odot}$ (right
panel) as obtained in a SIDM model.
Each halo have been modeled 
assuming the cross section value above (in the figure the corresponding
density profiles have empty circles) and a cross section three times 
larger (filled circles). In the same plot solid lines are the NFW
profiles.  
Within a Hubble time a modest self-interaction cross section value
creates soft cores in haloes of any size and a three times stronger  
cross section value  produces larger soft cores.
Bottom panels of Fig. 2 show the corresponding halo radial dispersion velocity
profiles (symbols have the same meanings as above). 
This is the case in which any trend towards the core collapse
is avoided by the competition between a mass accretion 
determined by the halo merging history and a thermalization process
by collisions.

\section{Conclusions}
\vskip -0.3truecm
H$\alpha$ rotation curves carried out at TNG of two late-type 
dwarf galaxies
 and one LSB galaxy show evidence of soft cores as  
the mass distribution of CL0024+1654 
on galaxy cluster scales.
Integrating the information from galaxies to clusters
of galaxies the halo central density seems to be independent of the
halo mass.
WDM in unable to produce visible soft cores, 
but SIDM in a {\it very weak} self-interacting regime 
($\sigma/m_x v_{100}=10^{-24}$ cm$^2$/Gev)  produces  
soft cores in agreement with the available observations on different scales.
\section{Acknowledgments}

E.D. and D.M. are grateful to Richard Bower and Oleg Gnedin for
stimulating discussions about this topic during the congress.
\section{References}
\verb''Avila-Reese V.,Colin P.,Valenzuela O.,D'Onghia E.,Firmani C.,
2001,(astro-ph/0010525)\\
\verb''Firmani C.,D'Onghia E.,Avila-Reese V.,Chincarini G.,
Hern\'{a}ndez X.,2000,MNRAS,315,L29\\
\verb''Moore B., Quinn T., Governato F., Stadel J., Lake G., 1999, MNRAS, 310,
 1147\\
\verb''Navarro J., Frenk C.S., White S.D.M., 1997, ApJ, 490, 493\\
\verb''Spergel D., Steinhardt P., 2000, PhRvL, 84, 3760\\
\verb''Tyson J.A., Kochanski G.P., Dell'Antonio I.P., 1998, ApJ, 498, L107\\
\verb''van den Bosch F.C., Swaters R.A., 2001, MNRAS, 325, 1017\\

\end{document}